\documentclass[english,11pt]{article}
\usepackage[T1]{fontenc}
\usepackage[utf8]{inputenc}
\usepackage{amsmath}
\usepackage{amssymb}
\usepackage{bm}
\usepackage{babel}
\RequirePackage[numbers,sort&compress]{natbib}
\usepackage[colorlinks=true,urlcolor=blue,linkcolor=blue,citecolor=red]{hyperref}

\def\be{\begin{equation}}
\def\ee{\end{equation}}
\def\ba{\begin{eqnarray}}
\def\ea{\end{eqnarray}}
\def\onehalf{{\textstyle{\frac{1}{2}}}}

\def\mt{{\mbox{\tiny{(1)}}}}

\begin{document}

\begin{center}
{\Large \bf de Sitter-invariant special relativity and}\\
\vskip 0.15cm 
{\Large \bf the dark energy problem}
\vskip 0.5cm
{\bf A. Araujo, D. F. L\'opez and J. G. Pereira}
\vskip 0.1cm
{\it Instituto de F\'{\i}sica Te\'orica, Universidade Estadual Paulista \\
Rua Dr. Bento Teobaldo Ferraz 271 \\
01140-070 S\~ao Paulo, Brazil}
\end{center}

\vskip 0.3cm
\centerline{\bf Abstract}
\begin{quote}
{\footnotesize The replacement of the Poincar\'e-invariant Einstein special relativity by a de Sitter-invariant special relativity produces concomitant changes in all relativistic theories, including general relativity. A crucial change in the latter is that both the background de Sitter curvature and the gravitational dynamical curvature turns out to be included in a single curvature tensor. This means that the cosmological term $\Lambda$ no longer explicitly appears in Einstein equation, and is consequently not restricted to be constant. In this paper, the Newtonian limit of such theory is obtained, and the ensuing Newtonian Friedmann equations are show to provide a good account of the dark energy content of the present-day universe.

}
\end{quote}

\section{Introduction}
\label{intro}

The de Sitter spacetime is usually interpreted as the simplest {\em dynamical} solution of the sourceless Einstein equation in the presence of a cosmological constant, standing on an equal footing with all other gravitational solutions, like for example Schwarzschild and Kerr. However, as a non-gravitational spacetime (in the sense that its metric does not depend on Newton's gravitational constant), the de Sitter solution should instead be interpreted as a fundamental background for the construction of physical theories, standing on an equal footing with the Minkowski solution. General relativity, for instance, can be constructed on any one of them. Of course, in either case gravitation will have the same dynamics, only their local kinematics will be different. If the underlying spacetime is Minkowski, the local kinematics will be ruled by the Poincar\'e group of special relativity. If the underlying spacetime is de Sitter, the local kinematics will be ruled by the de Sitter group, {\it which amounts then to replace ordinary special relativity by a de Sitter-invariant special relativity} \cite{dSsr0,dSsr1,livro}.\footnote{The first ideas about a de Sitter special relativity are due to L. Fantappi\'e, who in 1952 introduced what he called {\it Projective Relativity}, a theory that was further developed by G. Arcidiacono. The relevant literature can be traced back from Ref.~\cite{FA}.}

It turns out that there is a physical motivation for making such replacement. It is related to the existence of an invariant length parameter at the Planck scale, represented by the Planck length. The problem is that the Lorentz group is believed no to allow the existence of an invariant length parameter. Since Lorentz is a subgroup of Poincar\'e---which is the group that rules the kinematics in ordinary special relativity---the kinematics at Planck scale cannot be described by ordinary special relativity. This, however, does not mean that Lorentz symmetry must be violated at the Planck scale. To understand why, let us first recall that Lorentz transformations can be performed only in {\it homogeneous spacetimes}. In addition to Minkowski, therefore, they can be performed in de Sitter and anti-de Sitter spaces, which are the unique homogeneous spacetimes in $(1+3)$-dimensions \cite{W}. In what follows, our interest will be restricted to the de Sitter spacetime and group. 

As a homogeneous space, the de Sitter spacetime has constant sectional curvature. Of course, the Ricci scalar is also constant and has the form
\be
R = 12 \, l^{-2} \, ,
\label{RicciScalar}
\ee
where $l$ is the de Sitter length-parameter, or pseudo-radius. Now, by definition, Lorentz transformations do not change the curvature of the homogeneous spacetime in which they are performed. Since the scalar curvature is given by (\ref{RicciScalar}), {\it Lorentz transformations are found to leave the length parameter $l$ invariant} \cite{ccc}. Although somewhat hidden in Minkowski spacetime, because what is left invariant in this case is an infinite length---corresponding to a vanishing scalar curvature---in de Sitter spacetime, whose pseudo-radius is finite, this property becomes manifest. Contrary to the usual belief, therefore, {\it Lorentz transformations do leave invariant a very particular length parameter: that defining the scalar curvature of the homogeneous spacetime in which they are performed}. If the Planck length $l_P$ is to be invariant under Lorentz transformations, it must then represent the pseudo-radius of spacetime at the Planck scale, which will be a de Sitter space with a Planck cosmological term
\be
\Lambda_P = 3 / l_P^{2} \simeq 1.2 \times 10^{70}\, {\rm m}^{-2}.
\label{PlanLam}
\ee
In a de Sitter-invariant special relativity, therefore, the existence of an invariant length-parameter at the Planck scale does not clash with Lorentz invariance, which remains a symmetry at all scales. Taking into account the deep relationship between Lorentz symmetry and causality \cite{zeeman}, in this theory causality is always preserved, even at the Planck scale. Instead of Lorentz, translation invariance is broken down. In fact, in this theory, physics turns out to be invariant under the so-called {\it de Sitter translations}, which in stereographic coordinates are given by a combination of translations and proper conformal transformations \cite{dStrans}. We can then say that, in the same way Einstein special relativity may be thought of as a generalization of Galilei relativity for velocities near the speed of light, the de Sitter-invariant special relativity may be thought of as a generalization of Einstein special relativity for energies near the Planck energy. It holds, for this reason, at all energy scales.

When the Poincar\'e-invariant Einstein special relativity is replaced by a de Sitter-invariant special relativity, general relativity changes to what we have called {\it de Sitter-modified general relativity}.\footnote{It is opportune to remark that, even though both the de Sitter-modified general relativity and the MacDowell-Mansouri theory \cite{MDM} rely on the de Sitter symmetry, they are completely different theories. In fact, whereas the first is a modified gravity that presents a different phenomenology in relation to general relativity, the second is simply an alternative way to formulate general relativity.} In this theory, the kinematic curvature of the underlying de Sitter spacetime and the dynamical curvature of general relativity are both included in a single Riemann tensor. This means that the cosmological term $\Lambda$ no longer appears explicitly in Einstein's equation, and consequently the second Bianchi identity does not require it to be constant \cite{hendrik}. Far away from the Planck scale, $\Lambda$ can consequently assume smaller values, corresponding to larger values of the de Sitter length-parameter $l$. For low energy systems, like for example the present-day universe, the value of $\Lambda$ will be very small, and the de Sitter invariant special relativity will approach the Poincar\'e-invariant Einstein special relativity.

Spacetimes that do not reduce locally to Minkowski are known since long and come under the name of Cartan geometry \cite{CartanGeo}. The particular case in which it reduces locally to de Sitter is known in the literature as de Sitter-Cartan geometry~\cite{wise}. By considering general relativity in such geometry, we are going first to study its Newtonian limit. Then, by using this limit, we obtain the corresponding Newtonian Friedmann equations with basic purpose of examinig the ability of this theory to deal with cosmology, and in particular with the dark energy problem.

\section{The local value of the cosmological term $\Lambda$}

Implicit in de Sitter-modified general relativity is the idea that any physical system with energy density $\varepsilon_m$ induces a local cosmological term $\Lambda$ in spacetime, with an energy density $\varepsilon_\Lambda$, which is necessary to comply with the local symmetry of spacetime, now ruled by the de Sitter group.\footnote{The idea that the presence of matter with an energy density $\varepsilon_m$ could somehow change the underlying spacetime from Minkowski to de Sittter, was first put forward by F. Mansouri in 2002 \cite{mansouri}.} Note that the ensuing cosmological term $\Lambda$ is different from the usual notion in the sense that it is not constant. For example, outside the region occupied by the physical system, where $\varepsilon_m$ vanishes, $\Lambda$ vanishes as well. In a sense we can say that it represents an asymptotically flat local de Sitter spacetime. The question then arises: given a physical system, how to obtain the local value of the cosmological term? To answer this question, let us recall that, at the Planck scale, the cosmological term assumes the Planck value given by (\ref{PlanLam}). Since $\Lambda$ represents ultimately an energy density, this expression can be rewritten in the form
\be
\Lambda_P = \frac{4 \pi G}{c^4} \, \varepsilon_P  \, 
\label{kineLambda0}
\ee
where
\be
\varepsilon_P = \frac{m_P \, c^2}{(4 \pi/3) l_P^3}
\ee
is the Planck energy density, with $m_P$ the Planck mass. Now, the very definition of $\Lambda_P$ can be considered an extremal particular case of a general expression relating the local cosmological term to the corresponding energy density of a physical system. Accordingly, to a physical system of energy density $\varepsilon_m$ will be associated the local cosmological term
\be
\Lambda = \frac{4 \pi G}{c^4} \, \varepsilon_m \, .
\label{kineLambda}
\ee
As an example, let us consider the present-day universe. According to the results of WMAP, the space section of spacetime is nearly flat today. As a consequence, the mean energy density of the universe is of the same order of the critical energy density, which is equivalent to $\varepsilon_m \simeq 10^{-9}~{\rm Kg} \, {\rm m}^{-1} \, {\rm s}^{-2}$. Using this value, the effective cosmological term of the present-day universe is found to be
\be
\Lambda \simeq  10^{-52}~\mbox{m}^{-2} \, ,
\label{LambdaToday}
\ee
which is of the order of magnitude of the observed value \cite{obs1,obs2,obs3}.

It is important to remark that, although $\Lambda$ is allowed to change in space and time, it is not a dynamic, but a kinematic variable. This means that the {\it dynamical} field equations of the theory are unable to account for its space and time evolution. It is actually an external parameter, and as such assumed to be constant when solving the dynamical field equations of the theory. Once the field equations are solved, including their Newtonian limit, one can then use the values of $\Lambda$ obtained from the algebraic relation (\ref{kineLambda}). 

\section{Diffeomorphism in de Sitter-Cartan geometry}
\label{KillingStatic}

The de Sitter spacetime $dS$ can be seen as a hypersurface in the $(1+4)$-dimensional ``host'' pseudo-Euclidean space with metric
\[
\eta_{AB} = (+1,-1,-1,-1,-1),
\]
inclusion whose points in Cartesian coordinates $\chi^A = (\chi^\mu, \chi^4)$ satisfy \cite{ellis}
\be
\eta_{\mu \nu} \chi^\mu \chi^\nu - (\chi^4)^2 = - \, l^2 .
\ee
In terms of the host space coordinates $\chi^A$, an infinitesimal de Sitter transformation is written as
\be
\delta \chi^A = \onehalf \, \epsilon^{BC} \xi^{~A}_{BC} \, ,
\ee
where $\epsilon^{BC} = - \epsilon^{CB}$ are the transformation parameters, and
\be
\xi^{~A}_{BC} = \chi_{B} \, \delta^A_C - 
\chi_{C} \, \delta^A_B 
\label{KillingdS}
\ee
are the Killing vectors of the de Sitter group. The components
\be
\xi^{\;\alpha}_{\beta \gamma} = \chi_{\beta} \, \delta^\alpha_\gamma - 
\chi_\gamma \, \delta^\alpha_\beta
\label{Killing1}
\ee
represent the Killing vectors of the Lorentz group, whereas the components
\be
\xi^\alpha_\beta \equiv l^{-1} \xi^{\;\alpha}_{\beta 4} = 
l^{-1} \big(\chi_{4} \, \delta^\alpha_\beta - 
\chi_\beta \, \delta^\alpha_4 \big) = l^{-1}  \chi_{4} \, \delta^\alpha_\beta
\label{Killing2}
\ee
represent the Killing vectors of the de Sitter ``translations''.

Let us consider now the static coordinates $(ct, r, \theta, \varphi)$. They can be obtained from the embedding coordinates $\chi^A$ through the projection
\ba
\chi_0 &=& l \sqrt{1 - r^2/l^2} \, \sinh(ct/l) \label{p1} \\
\chi_1 &=& r \sin \theta \sin \varphi \label{p2} \\
\chi_2 &=& r \sin \theta \cos \varphi \label{p3} \\
\chi_3 &=& r \cos \theta \label{p4} \\
\chi_4 &=& l \sqrt{1 - r^2/l^2} \, \cosh(ct/l) \label{p5} \, .
\ea
The de Sitter metric in terms of the embedding coordinates is
\be
ds^2 \equiv \eta_{AB} \, d \chi^A d \chi^B =
(d \chi^0)^2 - (d \chi^1)^2 - (d \chi^2)^2 - (d \chi^3)^2 - (d \chi^4)^2 \; .
\ee
Using Eqs.~(\ref{p1}-\ref{p5}), one can easily verify that it is
\be
ds^2 = \big(1 - {r^2}/{l^2}\big) \, c^2 dt^2 - \frac{dr^2}{1 - {r^2}/{l^2}} - r^2 \left(d\theta^2 + \sin^2\theta \, d\varphi^2 \right).
\label{dSmetric}
\ee
Similarly, it is possible to obtain the Killing vectors of the de Sitter group in static coordinates. In particular, the Killing vectors (\ref{Killing2}) associated to the de Sitter ``translations'' are found to be
\be
\xi^\alpha_\mu = (1 - r^2/l^2)^{1/2} \cosh(ct/l) \, \delta^\alpha_\mu \, .
\label{pi}
\ee
The corresponding generators are then written in the form
\[
\Pi_\mu = \xi^\alpha_\mu \, \partial_\alpha \, .
\]

Now, as a local transformation, diffeomorphisms are able to detect the local structure of spacetime. In a locally-de Sitter spacetime, therefore, a diffeomorphism will be defined by
\be
\delta_{\Pi} x^\mu = \xi^{\mu}_{\alpha} \, \epsilon^{\alpha}(x) \, ,         
\label{dStrans}
\ee
where $\xi^{\mu}_{\alpha}$ are the Killing vectors (\ref{pi}) of the de Sitter ``translations''. In the contraction limit $l \to \infty$, the underlying de Sitter spacetime reduces to Minkowski \cite{gursey}, and the diffeomorphism (\ref{dStrans}) reduces to the diffeomorphism of locally-Minkowski spacetimes,
\be
\delta x^\mu = \delta^\mu_\alpha \, \epsilon^\alpha(x) \, ,
\label{OrDiff}
\ee
with $\delta^\mu_\alpha$ the Killing vectors of ordinary translations. 

\section{Einstein equation in locally-de Sitter spacetimes}
\label{modEHa}

To begin with, let us consider the action integral of a general source field
\be
{\mathcal S}_m = \frac{1}{c} \int {\mathcal L}_m \, \sqrt{-g} \, d^4x \, ,
\label{Am}
\ee
with ${\mathcal L}_m$ the lagrangian density. Invariance of this action under the diffeomorphism (\ref{dStrans}) yields, through Noether's theorem, the conservation law \cite{dSgeod}
\be
\nabla_\mu \Pi^{\rho \mu} = 0 \, ,
\label{UniCon}
\ee
where the conserved current has the form
\be
\Pi^{\rho \mu} = \xi^{\rho}_{\alpha} \, T^{\alpha \mu} \, ,
\label{EMtensor}
\ee
with $T^{\alpha \mu}$ the symmetric energy-momentum current. On the other hand, the Einstein-Hilbert action of general relativity in a locally-de Sitter spacetime is written as 
\be
{\mathcal S}_g = \int {R} \, \sqrt{-g} \, d^4x \, ,
\label{Sg}
\ee
with ${R}$ the scalar curvature obtained from
the de Sitter-Cartan curvature tensor ${R}^\alpha{}_{\beta \mu \nu}$, which is a tensor that represents both the dynamical curvature of general relativity and the kinematic curvature of the underlying de Sitter spacetime.

We consider now the total action integral
\be
{\mathcal S} = {\mathcal S}_g + {\mathcal S}_m \, .
\ee
The invariance of ${{\mathcal S}}$ under the diffeomorphism (\ref{dStrans}) yields the de Sitter-modified Einstein equation
\be
{R}_{\mu \nu} - {\onehalf} g_{\mu \nu} {R} = 
\frac{8 \pi G}{c^4} \Pi_{\mu \nu} \, .
\label{NewEinstein}
\ee
This is the equation that replaces ordinary Einstein equation when the Poincar\'e-invariant special relativity is replaced by a de Sitter-invariant special relativity. A crucial point of this approach is that, since both the dynamical curvature of general relativity and the kinematic curvature of the underlying de Sitter spacetime are now included in the Riemann tensor ${R}^\alpha{}_{\lambda \mu \nu}$, the (contracted form of the second) Bianchi identity,
\be
\nabla_\mu \big({R}^{\mu \nu} - 
{\onehalf} g^{\mu \nu} {R} \big) = 0 \, ,
\label{CBI}
\ee
does not require $\Lambda$ to be constant. It should be noted that such non-constant $\Lambda$ is possible at the expense of violating the conservation of the usual notions of energy and momentum \cite{EnerVio}. What is conserved now is the projection of the energy-momentum tensor along the Killing vectors associated to the de Sitter ``translations'', as given by Eq.~(\ref{EMtensor}).

\section{Linearization of the de Sitter-modified Einstein equation}

Let us rewrite the field equation (\ref{NewEinstein}) in the form
\be
{R}_{\mu \nu} = \frac{8 \pi G}{c^4} \Big( \Pi_{\mu \nu} -
\onehalf \, g_{\mu \nu} \Pi \Big) \, .
\label{EE}
\ee
In a de Sitter-Cartan geometry, in which the background spacetime is de Sitter instead of Minkowski, the spacetime metric is expanded in the form
\be
g_{\mu\nu} = \hat{g}_{\mu\nu} + h_{\mu\nu},
\ee
where $\hat{g}_{\mu\nu}$ represents the background de Sitter metric and $h_{\mu\nu}$ is the metric perturbation. The background connection, which corresponds to the zeroth-order connection, is
\be
\hat \Gamma^{\rho}{}_{\mu \nu} = \onehalf \, \hat{g}^{\rho \lambda}
\big(\partial_\mu \hat{g}_{\lambda \nu} + \partial_\nu \hat{g}_{\mu \lambda} -
\partial_\lambda \hat{g}_{\mu \nu} \big).
\label{0conn}
\ee
The corresponding Riemann tensor $\hat{R}^\alpha{}_{\beta \mu \nu}$ represents the curvature of the (non-gravitational) de Sitter background.

The first-order connection, on the other hand, is given by
\be
\Gamma_{\mt}^{\,\rho}{}_{\mu\nu} = \onehalf \,
\hat g^{\mu \nu} \big(\partial_{\mu} h^{\rho}{}_{\nu} +
\partial_{\nu} h^{\rho}{}_{\mu} - \partial^{\rho} h_{\nu\mu}  \big) - \onehalf \,
h^{\rho \lambda}
\big(\partial_\mu \hat{g}_{\lambda \nu} + \partial_\nu \hat{g}_{\mu \lambda} -
\partial_\lambda \hat{g}_{\mu \nu} \big).
\ee
After some algebraic manipulation, it can be rewritten in the form
\be
\Gamma_{\mt}^{\,\rho}{}_{\mu\nu} = \onehalf \big(\hat \nabla_{\mu}h^{\rho}{}_{\nu} +
\hat \nabla_{\nu} h^{\rho}{}_{\mu} - \hat \nabla^{\rho}h_{\nu\mu} \big),
\label{christoffel}
\ee
with $\hat \nabla_{\mu}$ a covariant derivative in the de Sitter connection (\ref{0conn}). The corresponding first-order Ricci tensor is
\be
R^\mt_{\mu\nu} = \onehalf \hat \nabla_{\rho} \hat \nabla_{\nu}h^{\rho}{}_{\mu} +
\onehalf \hat \nabla_{\rho} \hat \nabla_{\mu}h^{\rho}{}_{\nu} -
\onehalf \hat \nabla^{\rho} \hat \nabla_{\rho}h_{\mu\nu} -
\onehalf \hat \nabla_{\mu} \hat \nabla_{\nu}h,
\label{ricci}
\ee
where $h=h^{\rho}{}_{\rho}$. Using the identity\footnote{We use the same notation and conventions of Ref.~\cite{landau}.}
\be
\hat \nabla_\rho \hat \nabla_\mu h^\rho{}_\nu -
\hat \nabla_\mu \hat \nabla_\rho h^\rho{}_\nu =
h^\sigma{}_\nu \, \hat R_{\sigma \mu} -
h^\rho{}_\sigma \, \hat R^\sigma{}_{\nu \rho \mu} \, ,
\ee
we get
\be
R^\mt_{\mu\nu} = - \, \onehalf \hat \Box h_{\mu \nu} +
\onehalf \hat \nabla_\mu \big( \hat \nabla_\rho h^\rho{}_\nu - \onehalf \hat \nabla_\nu h \big) 
+ \onehalf \hat \nabla_\nu \big( \hat \nabla_\rho h^\rho{}_\mu - \onehalf \hat \nabla_\mu h \big) +
h^\sigma{}_{(\nu} \, \hat R_{\sigma \mu)} -
h^\rho{}_\sigma \, \hat R^\sigma{}_{(\mu \rho \nu)},
\label{Ricci1}
\ee
with the parentheses indicating a symmetrization in the neighbor indices.

At the first order, the class of harmonic coordinates is obtained by imposing the condition
\be
\hat{g}^{\mu\nu} \, {\Gamma}_\mt^{\rho}{}_{\mu\nu} = 0 \, ,
\ee
which can be recast in the form
\be
\hat \nabla_{\nu}h^{\rho\nu} - \onehalf \hat \nabla^{\rho}h = 0.
\label{armonico}
\ee
Using this condition in (\ref{Ricci1}), the first-order Ricci tensor is found to be
\be
R^\mt_{\mu\nu} = - \, \onehalf \hat \Box h_{\mu \nu} +
h^\sigma{}_{(\nu} \, \hat R_{\sigma \mu)} -
h^\rho{}_\sigma \, \hat R^\sigma{}_{(\mu \rho \nu)}.
\label{Ricci1bis}
\ee
At this order the de Sitter-modified Einstein equation (\ref{EE}) assumes then the form
\be
- \, \onehalf \hat \Box h_{\mu \nu} +
h^\sigma{}_{(\nu} \, \hat R_{\sigma \mu)} -
h^\rho{}_\sigma \, \hat R^\sigma{}_{(\mu \rho \nu)} = \frac{8 \pi G}{c^4} \Big( \Pi_{\mu \nu} -
\onehalf \, \hat g_{\mu \nu} \Pi \Big) \, .
\label{EEbis}
\ee
For the sake of comparison we recall that in ordinary general relativity, where spacetime reduces locally to Minkowski, the corresponding field equation has the form
\be
- \, \onehalf \Box h_{\mu \nu} = \frac{8 \pi G}{c^4} \Big( T_{\mu \nu} -
\onehalf \, \hat g_{\mu \nu} T \Big) \, .
\ee

\section{Newtonian limit}
\label{Nl}

The Newtonian limit is obtained when the gravitational field is weak and the particle velocities are small. In the presence of a cosmological term $\Lambda$, on the other hand, it has some subtleties related to the process of group contraction. Notice, to begin with, that the Galilei group is obtained from Poincar\'e under the contraction limit $c \to \infty$. The Newton-Hooke group, however, does not follow straightforwardly from the de Sitter group through the same limit. The reason is that, under such limit, the boost transformations are lost. In order to obtain a physically acceptable result, one has to simultaneously consider the limits $c \to \infty$ and $\Lambda \to 0$, but in such a way that
\be
\lim c^2 \Lambda = \frac{1}{\tau^2} \, 
\ee
with $\tau$ a time parameter. This means that the usual weak field condition of Newtonian gravity must be supplemented by the small $\Lambda$ condition \cite{gibbons}
\be
\Lambda r^2 \ll 1\, ,
\ee
which is equivalent to $r^2/l^2 \ll 1$. Accordingly, in what follows we will keep terms up to order $r^2/l^2$; terms of order $r/l^2$ will be discarded as they represent corrections to Newtonian limit.

\subsection{de Sitter-modified Poisson equation}

In the Newtonian limit, only the component $R^\mt_{00}$ is needed. In this case, the last term on the right-hand side of (\ref{Ricci1bis}) vanishes. Identifying furthermore
\be
h_{00} = 2 \phi/c^2,
\ee
with $\phi$ the gravitational scalar potential, we get
\be
R^\mt_{00} = \frac{2}{c^2} \Big[- \frac{1}{2} \, \hat \Box \phi +
\phi \, \hat R_{00} \Big] .
\label{Ricci00}
\ee
Neglecting the time derivatives in the d'Alembertian, we obtain
\be
R^\mt_{00} = \frac{1}{c^2} \Big[ \hat \Delta \phi +
2 \phi \, \hat R_{00} \Big] \, ,
\label{Ricci00b}
\ee
with $\hat \Delta$ the Laplacian in the de Sitter metric. Using this result in the de Sitter-modified Einstein equation (\ref{EEbis}), it becomes
\be
\hat \Delta \phi +
2 \phi \, \hat R_{00} = \frac{4 \pi G}{c^2} \, \Pi_{00} \, ,
\label{EE3}
\ee
where we have already used that $\Pi = \Pi^0{}_{0}$.

Now, in static coordinates, and up to the approximation we are using, the component $\hat R_{00}$ of the Ricci tensor is
\be
\hat R_{00} = - \, \frac{3}{l^2} \, (1 - r^2/l^2) \simeq - 3/l^2 \, ,
\ee
where we have discarded a term of order $r^2/l^4$. On the other hand, the source current is
\be
\Pi_{00} = \xi^0_0 \, T_{00} \, ,
\label{pixite}
\ee
with $\xi^0_0$ the zero-component of the Killing vectors of the de Sitter ``translations'', and $T_{00} = \rho c^2$. Substituting $\xi^0_0$ as given by Eq.~(\ref{k00}) of the Appendix, we get $\Pi_{00} = \rho_\Pi  c^2$, where
\be
\rho_\Pi \simeq \rho \left(1 - r^2/2 l^2 \right) \, .
\label{rhoPi}
\ee
Equation (\ref{EE3}) assumes then the form
\be
\hat \Delta \phi - \frac{6}{l^2} \phi = {4 \pi G} \rho_\Pi \, .
\label{EE4}
\ee

The Laplace operator $\hat \Delta$ in the background de Sitter metric $\hat{g}^{i j}$ is given by
\be
\hat \Delta \equiv \hat{g}^{i j} \hat{\nabla}_i \hat{\nabla}_j = 
\frac{1}{\sqrt{\hat{g}}} \, \partial_i \big(\sqrt{\hat{g}} \, \hat{g}^{ij} \partial_j \big)
\label{laplacian0}
\ee
with $\hat{g} = \det \hat{g}_{ij}$. Using the space components of the metric (\ref{dSmetric}), up to terms of order $r^2/l^2$, it is found to be
\be
\hat{\Delta} \phi = \frac{1}{r^2} \, \frac{\partial}{\partial r} 
\bigg(r^2 \, \frac{\partial \phi}{\partial r} \bigg) - \frac{r^2}{l^2} \frac{\partial^2 \phi}{\partial r^2} \, .
\label{laplacian1}
\ee
Equation (\ref{EE4}) can then be rewritten in the form
\be
\frac{1}{r^2} \, \frac{\partial}{\partial r} 
\bigg(r^2 \, \frac{\partial \phi}{\partial r} \bigg) - \frac{r^2}{l^2} \frac{\partial^2 \phi}{\partial r^2} - \frac{6}{l^2} \phi =
{4 \pi G} \rho_\pi \, .
\label{EE6}
\ee
The solution to this equation will be the de Sitter-modified Newtonian potential.

\subsection{The de Sitter-modified Newtonian potential}

In the contraction limit $l \to \infty$ (which corresponds to $\Lambda \to 0$), equation (\ref{EE6}) reduces to the usual Poisson equation
\be
\Delta \phi \equiv \frac{1}{r^2} \, \frac{\partial}{\partial r} 
\bigg(r^2 \, \frac{\partial \phi}{\partial r} \bigg)  = 4 \pi {G} \rho \, .
\label{poisson}
\ee
Its solution is given by
\be
\phi(r) = - \int \frac{G}{r - r'} \, \rho(r') \, dV' \, ,
\label{solutionfi}
\ee
where $r'$ is the distance from the volume element $dV'$ to the point where we are determining the potential. For a point particle located at ${\bm r}'$, the mass density is given by $\rho(r') = M \delta({\bm r} - {\bm r}')$ and we get
\be
\phi(r) = - \frac{G M}{r},
\label{Newton}
\ee
which is the Newtonian potential.

The same procedure should, in principle, be used to solve equation (\ref{EE6}). However, this is not necessary because, as an easy computation shows, if we replace
\be
\phi(r) \to \bigg(1 - \frac{r^2}{l^2} \bigg) \phi(r)
\ee
in the left-hand side of the ordinary Poisson equation (\ref{poisson}), up to terms of order $r/l^2$ it transforms into the left-hand side of the de Sitter-modified Poisson equation (\ref{EE6}). If $\phi(r)$ is solution of the ordinary Poisson equation (\ref{poisson}), the transformed potential $(1 - {r^2}/{l^2}) \phi(r)$ will be a solution of the de Sitter-modified Poisson equation (\ref{EE6}) with the same Green function:
\be
\bigg(1 - \frac{r^2}{l^2} \bigg) \phi(r) = - \int \frac{G}{r - r'} \, \rho_\Pi(r') \, dV' \, .
\label{solutionfiHat}
\ee
The solution can then be written in the form
\be
\phi(r) = - \bigg(1 - \frac{r^2}{l^2} \bigg)^{-1} \int \frac{G}{r - r'} \, 
\rho_\Pi(r') \,  dV' \, .
\label{solutionfiHatBis}
\ee
For a point particle located at ${\bm r}'$, the mass density $\rho_\Pi$ is given by
\be
\rho_\Pi(r') = M \bigg(1 - \frac{r'^2}{l^2} \bigg)^{\frac{1}{2}} \delta({\bm r} - {\bm r}') \, ,
\ee
and the solution is easily found to be
\be
\phi(r) = 
- \frac{G M}{r} - \frac{G M \Lambda}{6} \, r \, ,
\label{ModNewton1}
\ee
where we have used the relation $\Lambda = 3/l^2$. This is the de Sitter-modified Newtonian potential. The associated gravitational force $F = - \partial \phi/\partial r$ is
\be
F = - \frac{G M}{r^2} + \frac{G M \Lambda}{6} +
\frac{G M}{6} \, r \, \frac{\partial \Lambda}{\partial r} \, .
\label{ModNewtForceBis}
\ee
The first term represents the attractive Newtonian force. The de Sitter background contributes with an additional repulsive force, which is constant within regions where $\Lambda$ is constant. The last term is a new force that will be attractive or repulsive depending on whether $\partial \Lambda/\partial r$ is negative or positive. For physical systems in which $\Lambda$ is uniform, like for example the universe as a whole, this force vanishes, and (\ref{ModNewtForceBis}) reduces to
\be
F = - \frac{G M}{r^2} + \frac{G M \Lambda}{6} \, .
\label{ModNewtForceBis2}
\ee

It is opportune to mention here that, as we have already discussed, the underlying de Sitter spacetime changes the very notion of diffeomorphism, which in turn changes the form of the conserved current that appears in the right-hand side of Einstein equation. It so happens that such current appears behind the Newton gravitational constant $G$. This means that the de Sitter modification of Einstein equation, although kinematic in origin, acquires a gravitational character in the sense that it turns out to depend on the gravitational constant $G$, as can be seen from the $\Lambda$-term of the Newtonian force equation (\ref{ModNewtForceBis}). This should be compared with the expression of the Newton-Hooke force \cite{marek},
\be
F_{NH} = - \frac{GM}{r^2} + \frac{\Lambda c^2 r}{3} \, ,
\label{nh}
\ee
which is obtained in the Newtonian limit from the usual Einstein equation with a cosmological constant
\be
R_{\mu \nu} - \onehalf g_{\mu \nu} R - g_{\mu \nu} \Lambda =
\frac{8 \pi G}{c^4} T_{\mu \nu} \, .
\ee
In addition to a different dependence on the distance $r$, the $\Lambda$-part of the Newton-Hooke force (\ref{nh}) remains non-gravitational in the sense that it does not depend on Newton's gravitational constant.

\section{Newtonian Friedmann equations}

Newtonian cosmology was first discussed by Milne and McCrea in 1934 \cite{Mil,McMil}. The surprising result of this approach is that the Friedmann equations that follow from general relativity coincide with those obtained from Newtonian gravity. Of course, in spite of this coincidence, there are fundamental differences between the two approaches. For example, whereas in the Newtonian view the universe expands in flat Euclidian space under the influence of Newtonian gravity, in the relativistic view the whole universe consists of an expanding curved space. For many purposes, however, the Newtonian cosmology may still be used.\footnote{For a discussion of the properties and limitations of the Newtonian cosmology, see for example Ref.~\cite{Harrison}, Chapter 16.} Since our purpose here is not to study the time evolution of the universe, but just to explore the consequences of the de Sitter-invariant special relativity for the present-day universe, the Newtonian Friedmann equations should suffice.

Let us then begin by considering a sphere of radius ${\mathcal R} \equiv {\mathcal R}(t)$ and mass $M$ undergoing an isotropic and homogeneous expansion. The equation of motion for ${\mathcal R}$ can be obtained from the gravitational acceleration at the border of the sphere
\be
\frac{d^2{\mathcal R}}{dt^2} = - \frac{G M}{{\mathcal R}^2} + \frac{G M \Lambda}{6} \, ,
\ee
where we have used the de Sitter modified force (\ref{ModNewtForceBis2}), which is valid for the case of a uniform $\Lambda$. Multiplying both sides by $d{\mathcal R}/dt$ and integrating, we get the energy equation
\be
\frac{1}{2} \bigg( \frac{d {\mathcal R}}{dt}\bigg)^2 = \frac{G M}{{\mathcal R}} + 
\frac{G M \Lambda {\mathcal R}}{6} + E \, ,
\label{EnerEq1}
\ee
where the integration constant $E$ represents the total (kinetic plus potential) energy per unit mass at the surface of the expanding sphere. We write now the radius in the form
\be {\mathcal R}(t) = r \, a(t) \, ,
\ee
with $a(t) \equiv a$ the scale function parameter, and $r$ the co-moving radius of the sphere. Recalling that the mass of the sphere is
\be
M = \frac{4 \pi}{3} {\mathcal R}^3 \rho \, ,
\ee
with $\rho \equiv \rho(t)$ the mass density, after some algebraic manipulation, the energy equation (\ref{EnerEq1}) assumes the form
\be
\bigg(\frac{\dot a}{a} \bigg)^2 = \frac{8 \pi G}{3} \rho + 
\frac{4 \pi G \Lambda \rho {\mathcal R}^2}{9} + \frac{2 E}{{\mathcal R}^2} \, . 
\label{EnerEq2}
\ee

Now, in order to make contact with the Friedmann equations, the mass density $\rho$ must be replaced by the total density $\varepsilon_m/c^2$, where the subscript `$m$' denotes {\it all forms of matter energy, including the mass energy}. Furthermore, the energy $E$ must be related to the curvature of space. If we write
\be
E = - \frac{k c^2}{2}
\ee
with $k$ the curvature parameter, Eq.~(\ref{EnerEq2}) acquires the usual form of the Friedmann equations
\be
H^2 \equiv \bigg(\frac{\dot a}{a} \bigg)^2 = \frac{8 \pi G}{3 c^2} \, \varepsilon_m + 
\frac{4 \pi G \Lambda {\mathcal R}^2}{9 c^2} \, \varepsilon_m - 
\frac{k c^2}{{\mathcal R}^2} \, , 
\label{EnerEq3}
\ee
where $H = \dot a/a$ is the Hubble parameter. For a universe with a flat space section ($k = 0$), it becomes
\be
\varepsilon_c = \varepsilon_m + \frac{\Lambda {\mathcal R}^2}{6} \varepsilon_m \, ,
\label{67}
\ee
where $\varepsilon_c = {3 H^2 c^2}/{8 \pi G}$ is the Friedmann critical energy density. Since in this case $\varepsilon_c = \varepsilon_m + \varepsilon_\Lambda$, we can immediately identify the dark energy density as
\be
\varepsilon_\Lambda = \frac{\Lambda {\mathcal R}^2}{6} \, \varepsilon_m \, .
\label{DarkEnerDens}
\ee

We see from this expression that, according to the de Sitter-modified general relativity, any physical system with energy density $\varepsilon_m$ induces a local cosmological term $\Lambda$, with an associated dark energy density $\varepsilon_\Lambda$, which is necessary to comply with the local symmetry of spacetime, now ruled by the de Sitter group. We can then say that the presence of a local cosmological term $\Lambda$ is a natural consequence of the presence of ordinary matter. Most importantly, the dark energy density $\varepsilon_\Lambda$ is not a free parameter, but determined by the matter content of the universe through Eq.~(\ref{DarkEnerDens}). This means that, at each time of the universe evolution, one can infer the value of $\varepsilon_\Lambda$. For example, using the WMAP values for the current matter and dark energy density parameters
\be
\Omega_m \equiv \frac{\varepsilon_m}{\varepsilon_c} \simeq 0.28 \qquad {\rm and} 
\qquad \Omega_\Lambda \equiv \frac{\varepsilon_\Lambda}{\varepsilon_c} \simeq 0.72 \, ,
\label{wmap}
\ee
we obtain
\be
\varepsilon_\Lambda = 2.6 \; \varepsilon_m \, .
\label{DarkEnerDens2}
\ee
From Eq.~(\ref{DarkEnerDens}), this condition is satisfied, for example, if
\be
\Lambda \simeq 1.0 \times 10^{-52}~\mbox{m}^{-2} \qquad {\rm and} \qquad
{\mathcal R} \simeq 3.9 \times 10^{26}~{\rm m} \, ,
\ee
which are of the order of the observational values. Conversely, one can say that the observational values of $\Lambda$ and ${\mathcal R}$ yield the density parameters (\ref{wmap}). The present theory, therefore, provides a natural explanation for the so-called {\it coincidence problem}.


\section{Final remarks}
\label{Fr}

An appealing aspect of the theory presented in this paper is that it follows entirely from first-principles: one has just to replace the Poincar\'e-invariant Einstein special relativity by a de Sitter-invariant special relativity. Such replacement produces concomitant changes in all relativistic theories, including of course general relativity. According to this theory, dubbed de Sitter-modified general relativity, any physical system with energy-momentum density $\varepsilon_m$ induces in spacetime a local cosmological term (cf. Eq.~(\ref{kineLambda}))
\be
\Lambda = \frac{4 \pi G}{c^4} \, \varepsilon_m \, ,
\ee
which is necessary to comply with the local symmetry of spacetime, now ruled by the de Sitter group. In regions where no matter is present, $\varepsilon_m$ vanishes and $\Lambda$ vanishes as well. Notice that, differently from the usual notion of cosmological constant, the kinematic curvature of the background de Sitter spacetime and the dynamic curvature of general relativity are both included in the same Riemann tensor. As a consequence, the cosmological term is no longer restricted to be constant, being allowed to change in space and time. In addition, due to the change in the notion of diffeomorphism, the conserved currents change as well. For example, in static coordinates, the source Noether current has the form (cf. Eqs.~(\ref{EMtensor}) and (\ref{pi}))
\be
\Pi^{\mu \nu} \equiv \xi^\mu_\alpha \, T^{\alpha \nu} =
(1 - r^2/l^2)^{1/2} \cosh(ct/l) \; T^{\mu \nu} \, .
\label{dSGRsource}
\ee
The usual energy-momentum tensor, therefore, is no longer conserved. What is conserved now is the projection of the energy-momentum tensor along the Killing vectors associated to the de Sitter ``translations''.

To get a glimpse of how the de Sitter-modified general relativity works, let us consider first the solar system. Considering that no significant matter is present in the region between the Sun and the planets, the cosmological term $\Lambda$ is negligible in this region, and consequently no deviations are expected in relation to Newtonian gravity. Note that Earth, for example, produces a non-vanishing $\Lambda$ in the place where it is located. However, since its energy density is very small, $\Lambda$ will also be very small and no detectable effects are expected. On the other hand, for high energy photons, like for example those present in gamma ray bursts, the local value of $\Lambda$ can be large enough to interfere in the propagation of the photons themselves \cite{gamma}. Of course, since the energy of these photons are roughly fifteen orders of magnitude below the Planck energy \cite{GRB}, the interference will still be very small. However, considering that the physical effect of this interference is cumulative, and that the sources of gamma ray bursts are at very large distances from Earth, this effect could eventually be detectable. This is a crucial point in the sense that it could be used as an experimental test of the de Sitter-invariant special relativity.

For the universe as a whole, the matter energy density $\varepsilon_m$ gives rise to an effective cosmological term $\Lambda$, with an associated dark energy density $\varepsilon_\Lambda$. The dependence of $\varepsilon_\Lambda$ on  $\varepsilon_m$ establishes a relation between the matter density and the dark energy density, which in the Newtonian limit is given by Eq.~(\ref{DarkEnerDens}). When applied to the present-day universe, it gives a good account of the observed relation between $\varepsilon_\Lambda$ and $\varepsilon_m$, providing in this way an explanation for the coincidence problem. In addition, it is possible to envisage other properties of the universe. For example, considering that $\Lambda$ depends directly on the energy density of the universe, it might have assumed a huge value immediately after the big bang \cite{dSsrJGP}, which could account for inflation. Subsequently, it decayed together with the energy density of the universe, its current value being determined by the current energy density. Of course, in order to assess all properties of the theory, as well as the details of the ensuing cosmological model, the relativistic Friedmann equations for the de Sitter-modified Einstein equation should be obtained and studied. Based on the preliminary results obtained in this paper, such approach may constitute a new paradigm for the study of cosmology.

\begin{appendix}

\section*{Appendix: Newtonian limit of the Killing vectors}

As we have seen in Section~\ref{KillingStatic}, the Killing vectors associated to the de Sitter ``translations'' are
\be
\xi^\alpha_\mu = (1 - r^2/l^2)^{1/2} \cosh(ct/l) \, \delta^\alpha_\mu \, .
\label{TransKill}
\ee
As is well-known, the Newtonian limit is achieved by taking the limit $c \to \infty$. However, as we have discussed in Section~\ref{Nl}, in the presence of a cosmological term $\Lambda = 3/l^2$, such limit has physical meaning only if concomitantly we take $l \to \infty$, but in such a way that
\be
\lim_{c, l \to \infty} \frac{c}{l} = \frac{1}{\sqrt{3} \, \tau} \, ,
\ee
with $\tau$ a time parameter, sometimes interpreted as a cosmological time \cite{gibbons}. In this limit, therefore, the Killing vectors (\ref{TransKill}) assume the form
\be
\xi^\alpha_\mu = (1 - r^2/l^2)^{1/2} \cosh(t/\sqrt{3} \, \tau) \, \delta^\alpha_\mu \, ,
\label{TransKill2}
\ee
which no longer depends on $c$. Since the Killing vectors in the Newtonian limit must be static, we can choose $t$ such that $\cosh(t/\sqrt{3} \, \tau) = 1$, which yields
\be
\xi^\alpha_\mu = (1 - r^2/l^2)^{1/2} \, \delta^\alpha_\mu \simeq 
(1 - r^2/2 l^2)  \, \delta^\alpha_\mu \, .
\label{TransKill3}
\ee
These are the Newtonian Killing vectors. In particular, the component $\xi^0_0$ is
\be
\xi^0_0 \simeq (1 - r^2/2 l^2) \, .
\label{k00}
\ee
\end{appendix}

\section*{Acknowledgments}

AA and DFL would like to thank CAPES (33015015001P7) for fellowship grants. JGP thanks CNPq (309184/2013-4) for partial financial support.



\begin{thebibliography}{99}

\bibitem{dSsr0}
R. Aldrovandi, J. P. Beltr\'an Almeida and J. G. Pereira, {\it de Sitter special relativity}, Class. Quantum Grav. {\bf 24}, 1385 (2007); arXiv:gr-qc/0606122.

\bibitem{dSsr1}
S. Cacciatori, V. Gorini and A. Kamenshchik, {\it Special relativity in the $21^{\rm st}$ century}, {Ann. Phys. (Berlin)} {\bf 17}, 728 (2008);
arXiv:gr-qc/0807.3009.

\bibitem{livro}
R. Aldrovandi and J. G. Pereira, {\it An Introduction to Geometrical Physics}, second edition (World Scientific, Singapore, 2016), Chapter 37.

\bibitem{FA}
I. Licata and L. Chiatti, {\it The Archaic Universe: Big Bang, Cosmological Term, and the Quantum Origin of Time in Projective Cosmology}, Int. J. Theor. Phys. {\bf 22}, 48 (2009); arXiv:0808.1339.

\bibitem{W}
S. Weinberg, {\it Gravitation and Cosmology} (Wiley, New York, 1972).

\bibitem{ccc}
A. Araujo, H. Jennen, J. G. Pereira, A. C. Sampson and L. L. Savi, {\it On the spacetime connecting two aeons in conformal cyclic cosmology}, 
Gen. Rel. Grav. {\bf 47}, 151 (2015); arXiv:1503.05005.

\bibitem{zeeman}
E. C. Zeeman, {\it Causality implies Lorentz symmetry}, J. Math. Phys. {\bf 5}, 490 (1964).

\bibitem{dStrans}
R. Aldrovandi, J. P. Beltr\'an Almeida and J. G. Pereira, {\it A Singular conformal spacetime}, J. Geom. Phys. {\bf 56}, 1042 (2006), arXiv:gr-qc/0403099.

\bibitem{MDM}
S. W. MacDowell and F. Mansouri, {\it Unified Geometric Theory of Gravity and Supergravity}, Phys. Rev. Lett. {\bf 38}, 739 (1977).

\bibitem{hendrik}
H. Jennen, {\it Cartan geometry of spacetimes with non-constant cosmological function $\Lambda$}, Phys. Rev. D {\bf 90}, 084046 (2014); arXiv:1406.2621.

\bibitem{CartanGeo}
R. Sharpe, {\it Differential Geometry: Cartan's Generalization of Klein's Erlangen Program} (Springer, Berlin, 1997).

\bibitem{wise}
D. K. Wise, {\it MacDowell-Mansouri gravity and Cartan geometry}, Class. Quantum Grav. {\bf 27}, 155010 (2010); arXiv:gr-qc/0611154.
	
\bibitem{mansouri}
F. Mansouri, Phys. Lett. B, {\it Non-Vanishing Cosmological Constant, Phase Transitions,
and $\Lambda$--Dependence of High Energy Processes}, {\bf 538}, 239 (2002); arXiv:hep-th/0203150.

\bibitem{obs1}
A. G. Riess {\it et al}, {\it Observational Evidence from Supernovae for an Accelerating Universe and a Cosmological Constant}, Ap. J. {\bf 116}, 1009 (1998); arXiv:astro-ph/9805201.

\bibitem{obs2}
S. Perlmutter {\it et al}, {\it Measurements of Omega and Lambda from 42 High-Redshift Supernovae}, Ap. J. {\bf 517}, 565 (1999); arXiv:astro-ph/9812133.

\bibitem{obs3}
P. de Bernardis {\it et al}, {\it A Flat Universe from High-Resolution Maps of the Cosmic Microwave Background Radiation}, Nature {\bf 404}, 955 (2000); arXiv:astro-ph/0004404.

\bibitem{ellis}
S. W. Hawking and G. F. R. Ellis, {\it The Large Scale Structure 
	of Space-Time} (Cambridge University Press, Cambridge, 1973).

\bibitem{gursey}
F. G\"ursey, {\it Introduction to the de Sitter group}, in {\it Group Theoretical Concepts and Methods in Elementary Particle Physics}, ed. by F. G\"ursey (Gordon and Breach, New York, 1962).

\bibitem{dSgeod}
J. G. Pereira and A. C. Sampson, {\it de Sitter geodesics: reappraising the notion of motion}, Gen. Rel. Grav. {\bf 44}, 1299 (2012); arXiv:1110.0965.

\bibitem{EnerVio}
Thibaut Josset, Alejandro Perez, and Daniel Sudarsky, {\it Dark Energy from Violation of Energy Conservation}, Phys. Rev. Lett. {\bf 118}, 021102 (2017); arXiv:1604.04183. 

\bibitem{landau}
L. D. Landau and E. M. Lifshitz, {\it The Classical Theory of Fields}
(Butterworth-Heinemann, Oxford, 1980).

\bibitem{gibbons}
G. W. Gibbons and C. E. Patricot, {\it Newton-Hooke spacetimes, Hpp-waves and the cosmological constant}, Class. Quantum Grav. {\bf 20}, 5225 (2003); arXiv:hep-th/0308200.

\bibitem{marek}
See, for example, M. Nowakowski, {\it The Consistent Newtonian Limit of Einstein's Gravity with a Cosmological Constant}, Int. J. Mod. Phys. D {\bf 10}, 649 (2001);
arXiv:gr-qc/0004037.

\bibitem{Mil}
E. Milne, {\it A Newtonian expanding universe}, Quarterly J. Math. {\bf 5}, 64 (1934).

\bibitem{McMil}
W. H. McCrea and E. Milne, {\it Newtonian universes and the curvature of space}, Quarterly J. Math. {\bf 5}, 73 (1934).

\bibitem{Harrison}
E. Harrison, {\it Cosmology}, (Cambridge University Press, Cambridge, 2001).

\bibitem{gamma}
R. Aldrovandi and J. G. Pereira, {\it de Sitter Relativity: a New Road to Quantum Gravity?} 
Found. Phys. {\bf 39}, 1 (2009); arXiv:0711.2274.

\bibitem{GRB}
J. Albert {\it et al} (for the MAGIC Collaboration), J. Ellis, N. E. Mavromatos, D. V. Nanopoulos, A. S. Sakharov and E. K. G. Sarkisyan, {\it Probing quantum gravity using photons from a flare of the active galactic nucleus Markarian 501 observed by the MAGIC telescope}, Phys. Lett. B {\bf 668}, 253 (2008); arXiv/0708.2889.

\bibitem{dSsrJGP}
R. Aldrovandi, J. P. Beltr\'an Almeida and J. G. Pereira, {\it A Singular conformal spacetime}, J. Geom. Phys. {\bf 56}, 1042 (2006); arXiv:gr-qc/0403099.

\end{thebibliography}
\end{document}